\documentclass{article} 

\usepackage{graphicx}
\usepackage{physics}
\usepackage{hyperref}
\usepackage{amsmath}
\usepackage{setspace}
\usepackage{physics}
\usepackage{amssymb}
\usepackage{color}
\usepackage{url}
\usepackage{epstopdf}
\usepackage{authblk}

\title{Two Notions of Naturalness}
\author{Porter Williams
\thanks{Department of Philosophy, University of Southern California, Los Angeles, CA 90089}}
\date{}                     
\begin{document}
\maketitle

\begin{abstract}
My aim in this paper is twofold: (i) to distinguish two notions of naturalness employed in Beyond the Standard Model (BSM) physics and (ii) to argue that recognizing this distinction has methodological consequences. One notion of naturalness is an ``autonomy of scales'' requirement: it prohibits sensitive dependence of an effective field theory's low-energy observables on precise specification of the theory's description of cutoff-scale physics. I will argue that considerations from the general structure of effective field theory provide justification for the role this notion of naturalness has played in BSM model construction. A second, distinct notion construes naturalness as a statistical principle requiring that the values of the parameters in an effective field theory be ``likely'' given some appropriately chosen measure on some appropriately circumscribed space of models. I argue that these two notions are historically and conceptually related but are motivated by distinct theoretical considerations and admit of distinct kinds of solution.
\end{abstract}

\section{Introduction}

Since the late 1970s, attempting to satisfy a principle of ``naturalness'' has been an influential guide for particle physicists engaged in constructing speculative models of Beyond the Standard Model (BSM) physics. This principle has both been used as a constraint on the properties that models of BSM physics must possess and shaped expectations about the energy scales at which BSM physics will be detected by experiments. The most pressing problem of naturalness in the Standard Model is the Hierarchy Problem: the problem of maintaining a scale of electroweak symmetry breaking (EWSB) many orders of magnitude lower than the scale at which physics not included in the Standard Model becomes important.\footnote{One sometimes calls this problem the ``little Hierarchy problem,'' reserving the title ``Hierarchy problem'' for the problem of the hierarchy between the EWSB scale and the Planck scale. Nothing hinges on this distinction in this paper.} Models that provided natural solutions to the Hierarchy Problem predicted BSM physics at energy scales that would be probed by the LHC, and many particle physicists expected BSM physics to be detected. These expectations have been dashed by the first two runs of the LHC. The LHC is now probing energy scales above those at which many natural BSM models predicted new physics, but no new physics has been detected. This has led many in the particle physics community to reflect on what, precisely, the conceptual content of the naturalness principle is and what continuing role, if any, it ought to play in guiding current and future theorizing about BSM physics. One can usefully group the prevalent positions on the topic into three categories.\footnote{I neglect here the position that advocates simply staying the course, preserving the emphasis that has been placed on naturalness at the expense of focusing on more complicated natural extensions of the Standard Model.}


\begin{enumerate}

\item The Standard Model is unnatural and naturalness should play no role in the future of high-energy physics theorizing. 

\vspace{\baselineskip}

Advocates of this position argue that not only have the recent LHC results demonstrated that there is no natural explanation of the stability of the EWSB scale, we were mistaken to have thought there was any property of the EWSB scale crying out for explanation in the first place: the naturalness principle was ill-motivated from its inception. There have long been pockets of skepticism about the motivation for and influence of the naturalness principle; particularly trenchant expressions of such skepticism can be found in \cite{Richter2006} and, for post-LHC perspectives, \cite{Hossenfelder2018} and especially \cite{rosalerharlander2018}. It is also the case that even though many in the particle physics community have adopted naturalness as an important guide to model construction, many of the very same members of that community have also long been willing to seriously entertain certain unnatural models of BSM physics. This is illustrated by the influence of models with a high scale of supersymmetry breaking, such as models with ``split supersymmetry'', and related models \cite{Wells2003,GiudiceRomanino2004,Wells2005,ArkaniHamed2005}. 

\vspace{\baselineskip}

\item The Standard Model is unnatural but future BSM models should explain the EWSB scale in a way that also explains why no natural explanation was forthcoming.

\vspace{\baselineskip}

Advocates often accept that the naturalness principle was well-motivated within the context of effective field theory and that we were justified in seeking a natural explanation of the EWSB. However, they also accept that recent LHC results demonstrate that no natural explanation of the EWSB is likely and that a different, ``unnatural'' explanation should be sought. For some, this is a role that a multiverse can play in particle physics. For example, Giudice states that ``It is conceivable that the LHC will find that the Higgs mass does not respect the naturalness criterion, just like (probably) the case of the cosmological constant. Accepting this possibility, however, does not imply that we can simply ignore the issue\ldots if we accept Unnaturalness, we have to address the question of why the Higgs is unnatural. At the moment, the multiverse offers the most plausible answer at our disposal'' \cite[p. 4]{Giudice2013}. According to this position, turning to some version of a multiverse to explain the EWSB amounts to giving up on finding a natural explanation of the stability of the EWSB.\footnote{This position is also described in \cite{Weinberg2007,Barbieri2013electroWeak,Arvanitaki2014}, among others.}

\vspace{\baselineskip}

\item A multiverse provides a novel setting in which we can provide a \emph{natural solution} to the Hierarchy Problem and other problems of naturalness. 

\vspace{\baselineskip}

While the second position argues that pursuing a statistical explanation of the EWSB scale in a multiverse amounts to giving up on naturalness, this third position sees the multiverse as offering the possibility of what particle physicists have sought for decades: a natural solution to the Hierarchy problem and other problems of naturalness. This position relies on a particular, statistical notion of naturalness. According to this notion a property of a theory is considered natural if and only if it is ``likely'' or ``not improbable'' according to some chosen probability distribution.

\end{enumerate}

\noindent It is this third stance that receives extended critical scrutiny in this paper. 

The plan of the paper is as follows. In section 2, after a selective review of two episodes from the ``pre-history'' of naturalness, I present a condensed version of an argument I have made elsewhere \cite{Williams2015} that the best way to understand the content of the naturalness principle is as a prohibition of sensitive dependence of low-energy measurable quantities on comparatively high-energy physics. I argue this understanding renders naturalness arguments well-motivated within an effective field theory context and provides a single notion on which one can ground several apparently distinct formulations of naturalness in the physics literature.  In section 3, I briefly review the development of an alternative, statistical notion of naturalness that began in the 1990s. This development has made possible the recent proposal that naturalness problems could be embedded and given natural solutions within a multiverse. In section 4, I examine this proposal in greater detail and argue that the statistical notion of naturalness that it employs has little to do with the notion of naturalness that can actually be motivated by considerations drawn from effective field theory. I conclude with some brief remarks further distinguishing these two notions of naturalness and suggest that recognizing this distinction undermines one recently suggested source of evidential support for a multiverse.

\section{Naturalness and the autonomy of scales}

I want to begin by briefly describing two important episodes from 20th century physics in which different problematic aspects of elementary scalar particles were highlighted. These are the discovery by Weisskopf that the self-energy contribution to the mass of elementary scalar particles is quadratically divergent, and the much later recognition by Wilson that particle masses do not receive large radiative corrections if their mass terms are ``protected'' by a symmetry. My interest in these two episodes is decidedly Whiggish: I describe them here because both of these aspects were later incorporated into the notion of naturalness as distinct ways of describing what it is that makes elementary scalars ``unnatural''.

\subsection{A brief pre-history}

The recognition that there is something uniquely problematic about quantum field theories that contain elementary scalar particles goes back at least to Weisskopf \cite{Weisskopf1939}. Weisskopf was investigating the self-energy contribution to the mass of the electron and found that while the self-energy of the electron was logarithmically divergent, the self-energy for a charged scalar particle diverged quadratically.\footnote{Like many such calculations in the 1930s, Weisskopf made use of the hole-theoretic formalism of Dirac; see \cite[ch. 2]{Schweber1994} for calculational methods in the 1930s.} With these results in hand, Weisskopf ventured ``A few remarks\ldots about the possible significance of the logarithmic divergence of the self-energy for the theory of the electron'' [p. 75]. He begins his remarks by expanding the self-energy term, denoted $W$, in powers of $\frac{e^2}{hc}$, yielding the series expansion

$$W = \sum_n W_n = \sum_n c_n mc^2 \left(\frac{e^2}{hc}\right)^n\left[\ln\left(\frac{h}{mca}\right)^k\right] \,\, k \leq n$$

\noindent with $c_n$ unspecified constants and $a$ a length scale introduced to keep the logarithm finite. Weisskopf expresses the hope that this sum will converge if 

$$\delta = \left(\frac{e^2}{hc}\right)\left[\ln\left(\frac{h}{mca}\right)\right] < 1$$

\noindent in which case the self-energy contribution to the electron mass would be simply $\displaystyle W = mc^2 \mathcal{O}(\delta)$: the electron mass multiplied by a term of order $\delta$. 

On the assumption that the self-energy contribution is in fact given by $\displaystyle W = mc^2 \mathcal{O}(\delta)$, Weisskopf then aims to define an analogue of the classical electron radius by setting the self-energy $W = mc^2$. This is satisfied if the analogue of the classical electron radius, the ``critical length'', is 

$$\displaystyle a \sim \frac{h}{mc} \cdot \exp \left(\frac{-hc}{e^2}\right)$$ 

\noindent The significance of the critical length, according to Weisskopf, is that it indicates the length scale at which a theory becomes inconsistent and new physics must be present.\footnote{Weisskopf is here considering a theory to be inconsistent if the series expansion of the self-energy contribution $W$ does not converge.}

Turning his attention to a quantum field theory of elementary charged scalar particles, Weisskopf notes that ``the situation is entirely different'' than the case of the electron: the self-energy contribution to the mass of the scalar particle diverges quadratically. Using the same method as above to establish a critical length for a theory containing elementary charged scalar particles, Weisskopf determines that the critical length must be much larger than for the theory containing only electrons and positrons:

$$\displaystyle a \sim \left(\frac{hc}{e^2}\right)^{-\frac{1}{2}}\left(\frac{h}{mc}\right)$$

\noindent Weisskopf concludes his remarks as follows: 

\begin{quote}
This may indicate that a theory of particles obeying Bose statistics must involve new features at this critical length, or at energies corresponding to this length; whereas a theory of particles obeying the exclusion principle is probably consistent down to much smaller lengths or up to much higher energies \cite[p. 75]{Weisskopf1939}. 
\end{quote}

It is interesting that in addition to being the first to recognize that the self-energy contribution to the mass of an elementary scalar particle is quadratically divergent, Weisskopf also drew upon those calculations to extract from the breakdown of the scalar theory at a given length scale a prediction about the length scale at which new physics could be expected to exist: the scale $a$ such that for lengths shorter than $a$, Weisskopf conjectured that the series expansion of the self-energy was divergent.\footnote{This is not to suggest that Weisskopf's treatment of the scalar theory is unique in this regard; as shown above, he makes a similar argument in the case of the purely fermionic theory.} It is also worth noting that although the more modern understanding of quantum field theory in which the concept of naturalness developed was quite different from Weisskopf's own, it is still extremely common to find elementary scalar particles described as unnatural precisely because their mass terms are quadratically divergent.\footnote{For example, see \cite{Peskin1995,Murayama2000Supersymmetry,baertata2006,terning2006susy,Banks2008,Zee2010}.} Indeed, Zee refers to the quadratic divergence of elementary scalar particle masses, which he equates with the Hierarchy problem, as the ``Weisskopf phenomenon'' \cite[p. 419]{Zee2010}.


The second pre-historical episode is Wilson's recognition in \cite{Wilson1971} that elementary scalar particle mass terms diverge more severely than mass terms of fermions or gauge bosons because the scalar mass terms do not break any internal symmetries of a quantum field theory. This recognition comes in a paper in which Wilson was interested in applying the renormalization group (RG) methods of Gell-Mann and Low \cite{Gellmann1954} to the strong interactions; it is worth providing a brief review of the paper in order to offer some context for Wilson's remark. 

Wilson analyzes the behavior of the RG equation

$$\dv{\ln(\lambda)}\text{g}(\lambda) = \beta(\text{g}(\lambda))$$

\noindent for the case of a hypothetical theory containing a single scale-dependent coupling $\text{g}(\lambda)$. He approaches the RG equation as the equation of motion for a general dynamical system and focuses on its possible asymptotic behavior in the infrared (IR) ($\lambda\to 0$) and the ultraviolet (UV) ($\lambda\to\infty$) regions. He allows that the asymptotic behavior of the RG may be (i) a fixed point, i.e. the coupling $\text{g}(\lambda)$ hits a fixed point $g^*$ at which $\beta(\text{g}^*) = 0$, or (ii) a limit cycle, in which case the beta function approaches a periodic function with the period given by a function of the scale $\lambda$ and the values of the couplings oscillate perpetually (this situation requires extending the RG analysis to theories with more than one coupling). Only situation (i) is relevant for our purposes and I will not discuss situation (ii).

In examining the possible asymptotic fixed point solutions of the RG, Wilson further distinguishes between two physical possibilities: (a) it may be the case that weak and electromagnetic corrections to strong interaction processes remain small up to arbitrarily high energies (i.e. the weak and electromagnetic couplings remain small relative to the strong interaction coupling), in which case there is a theory that treats only the strong interactions that remains at least approximately valid for arbitrarily high momentum processes; or (b) there is an energy scale $\Lambda$ at which the weak and electromagnetic corrections become too large to be treated perturbatively (i.e. the weak and electromagnetic couplings become comparable in size to the strong coupling), which means that any theory that treats the strong interactions in isolation is valid only for momenta $\lambda\ll\Lambda$. 

Wilson begins by assuming that possibility (a) obtains and gives a more detailed, largely qualitative analysis of the possible fixed points of the RG equation for the coupling $\text{g}(\lambda)$.\footnote{See \cite[pp. 1825-6]{Wilson1971} for the assumptions about the behavior of the $\beta$-function underlying his analysis. He notes that ``there is no way of knowing whether these assumptions are true for quantum electrodynamics or any other given field theory.''} He asks the reader to imagine that the beta function has multiple asymptotic fixed points in the IR and the UV; the allowed IR fixed points are $\text{g}(0) = 0$ and $\text{g}(0) = x_2$, while the allowed UV fixed points are labeled $\text{g}(\infty) = x_1$ and $\text{g}(\infty) = x_3$. Wilson then partitions the possible values of $\text{g}(\lambda)$ into basins of attraction. He considers first the basins of attraction for the UV fixed points: initial values of the coupling $\text{g}(\lambda)$ in the range $0 < \text{g}(\lambda) < x_2$ all flow to the value $x_1$ as $\lambda\to\infty$, while values in the range $x_2 < \text{g}(\lambda) < x_3$ will flow to $x_3$ as $\lambda\to\infty$. 

\begin{center}

\begin{figure}
	\caption{Reproduced from \cite{Wilson1971}.}
  \includegraphics[keepaspectratio,width=\textwidth,height=8cm]{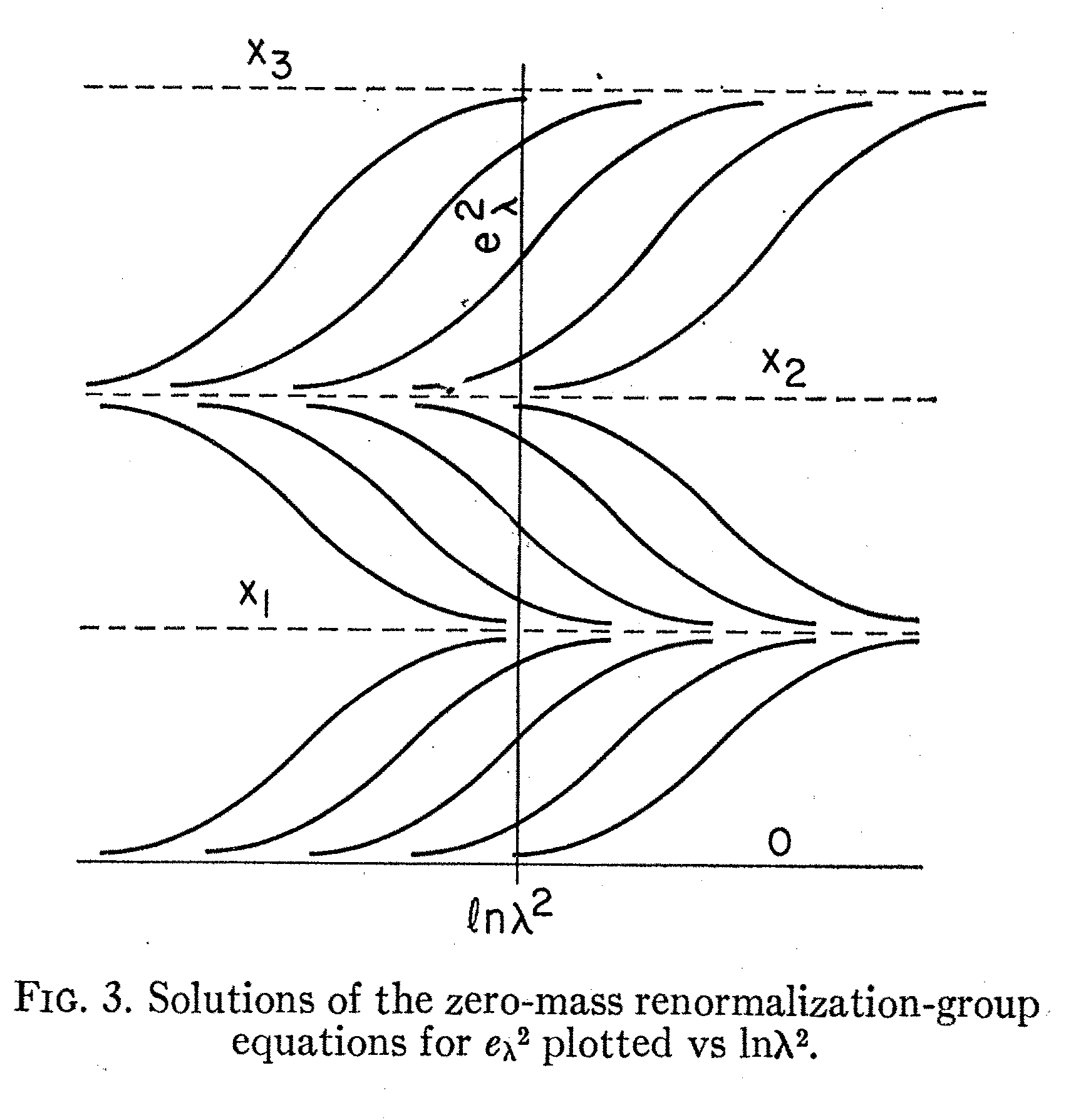}
\end{figure}

\end{center}

\noindent Turning to the IR fixed points, initial values of $\text{g}(\lambda)$ that lie in the range $0 < \text{g}(\lambda) < x_1$ flow to 0 as $\lambda\to 0$, while initial values $x_1 < \text{g}(\lambda) < x_3$ flow to $x_2$ as $\lambda\to 0$.

However, Wilson takes his analysis of the asymptotic UV behavior of the RG equation to suggest that the more plausible physical situation is possibility (b): there exists a UV cutoff scale $\Lambda\gg 1$ GeV at which any theory that treats the strong interactions in isolation necessarily breaks down. To make this argument, Wilson turns from a hypothetical theory of the strong interactions to quantum electrodynamics, while imagining that the analysis of the asymptotic UV behavior of the RG equations represented by Fig. 1 above holds for both cases.\footnote{Of course, we now recognize this was a mistake. Quantum chromodynamics, the theory of the strong interaction, is asymptotically free, while quantum electrodynamics is not.} 

Wilson first points out that the validity of perturbation theory in familiar applications of quantum electrodynamics justifies the assumption that the the initial value of the scale-dependent coupling $\text{g}(\lambda)$ is small for $\lambda\ll\Lambda$, i.e. sufficiently close to 0 that it lies in the range $0<\text{g}(\lambda)<x_2$. He then argues that his analysis illustrates that the coupling in quantum electrodynamics $\text{g}(\lambda)$ flows to the fixed-point value $x_1$ as $\lambda\to\infty$, and that this fixed-point value is independent of the value of $\text{g}(\lambda)$ at low energies.\footnote{In terminology that is now familiar, but which Wilson introduces immediately prior to this discussion, this is just to say that $x_1$ is an ultraviolet stable fixed point.} He concludes that ``this suggests that all particles will couple strongly to photons at sufficiently high momenta; but this would mean that electrodynamics and strong interactions would mix strongly, suggesting that pure electrodynamics is valid only below a cutoff momentum $\Lambda$'' \cite[p. 1832]{Wilson1971}. Referring back this analysis, Wilson reiterates this at the outset of the discussion in which he makes his remark about the mass terms of elementary scalar particles:\footnote{Wilson also cites \cite{Gellmann1954} and \cite{Bogoliubov1959} in support of the quotation below.}

\begin{quote}
Analysis of the renormalization group for electrodynamics shows that the $\lambda$-dependent charge $\text{e}_\lambda$ increases with $\lambda$, eventually becoming of order 1. By this is meant that no matter how small the renormalized charge $\text{e}$ is, $\text{e}_\lambda$ becomes of order some fixed number independent of $\text{e}$ if $\lambda$ is large enough.\footnote{Wilson has elsewhere defined the ``renormalized charge'' as the charge renormalized at $\lambda = 0$.} This suggests that there is a cutoff $\Lambda$ beyond which radiative corrections to strong interactions are too large to be treated as a perturbation. So it will be assumed here that the theory of strong interactions in isolation is valid only below the cutoff $\Lambda$ \cite[p. 1838]{Wilson1971}.
\end{quote}

\noindent Wilson thus treats possiblity (b) as the more physically likely one and explores what an RG analysis can teach us about the behavior of the strong interaction at energies $\text{E}\ll\Lambda$. 

Wilson also recognizes that the existence of such a UV cutoff would invalidate his above analysis of the UV fixed point structure. The reason is that his analysis was based on the assumption that the strong interactions could be treated in isolation up to arbitrarily high energy -- i.e. on the assumption of possiblity (a) -- and if possibility (b) obtains that is no longer the case. Instead, Wilson assumes that physical processes at energies higher than the cutoff are governed by a more complicated theory in which the strong, electromagnetic, and weak interactions become unified; it is this theory that determines the value of the strong coupling at the cutoff scale $\text{g}(\lambda=\Lambda)$. This means that on the one hand, the RG equation for $\text{g}(\lambda)$ will receive large weak and electromagnetic corrections as $\lambda$ approaches the cutoff $\Lambda$, since the weak and electromagnetic couplings become comparable in size to the strong coupling for $\lambda\sim\Lambda$. On the other hand, there is no reason to assume that as $\lambda\to\infty$ the asymptotic RG behavior of the more complicated Grand Unified Theory (GUT) that takes over above the cutoff scale $\Lambda$ will be similar to the asymptotic RG behavior of the theory that treated the strong interactions in isolation. Accordingly, Wilson turns his focus to the asymptotic IR behavior of the RG equation for $\text{g}(\lambda)$. 

It is at this point that Wilson makes a remark about a property that is unique to the mass terms of elementary scalar particles. He begins by noting that RG transformations leave the internal symmetries of a quantum field theory unbroken: if the set of couplings $\text{h}_{n\lambda}$ that would break the symmetry are zero at any value of $\lambda$ then these couplings remain zero at \emph{all} values of $\lambda$. This requires that any radiative corrections to a symmetry-breaking coupling $\text{h}_{n\lambda}$ must be proportional to the parameter $\text{h}_{n\lambda}$ itself. This ensures that when $\text{h}_{n\lambda} = 0$, 
any possible radiative corrections are also zero and thus that changing the scale at which the coupling is defined from $\lambda\to\lambda'$ will not break any internal symmetries of the theory. This is what it means for a coupling to be ``protected'' by a symmetry: it is ``protected'' from having its value altered significantly by radiative corrections. In particular, it is ``protected'' from acquiring a non-zero value purely from radiative corrections. 

Wilson then notes that to account for symmetries that are very weakly broken at high energies $\lambda\sim\Lambda$ but strongly broken at lower energies $\lambda\sim 1$ GeV, it must be the case that some symmetry-breaking couplings are very small at $\lambda\sim\Lambda$ but grow to $\mathcal{O}(1)$ as $\lambda\to 1$ GeV. This is only possible if those couplings do not receive large radiative corrections when $\lambda\sim\Lambda$, and that requires that those couplings be ``protected'' by a symmetry. In particular, Wilson says that if a quantum field theory describes light elementary particles, their mass terms must be ``protected'' by some symmetry:

\begin{quote}
\ldots all generalized mass terms must break an internal symmetry. A generalized mass term is any coupling which causes particles to have finite mass rather than zero mass. It is interesting to note that there are no weakly coupled scalar particles in nature; scalar particles are the only kind of free particles whose mass term does not break either an internal or a gauge symmetry.

This discussion can be summarized by saying that mass or symmetry-breaking terms must be ``protected'' from large corrections at large momenta due to various interactions (electromagnetic, weak, or strong)\ldots The mass terms for the electron and muon and the weak boson, if any, must also be protected. \emph{This requirement means that weak interactions cannot be mediated by scalar particles} \cite[p. 1840, emphasis added]{Wilson1971}.
\end{quote}

\noindent After stating that the weak interactions cannot be mediated by scalar particles, Wilson states that ``this rules out'' two such models: the models of \cite{kummersegre1965} and \cite{christ1968}.

Following 't Hooft \cite{hooft1980}, it is now common to hear that the fact that mass terms of elementary scalar particles are not ``protected'' by any symmetry makes those masses unnatural; it is noteworthy that Wilson recognized that this ``requirement'' was not met by elementary scalar masses already in 1971. More interesting is that he put this requirement to use as a criterion of theory selection, declaring that it rendered unviable quantum field theories in which the weak interactions were mediated by elementary scalar particles. This, too, is a familiar feature of the role that naturalness has played in more modern discussions.

Let me conclude this pre-historical discussion with a few remarks. First, I have not meant to suggest that these were the only pre-historical episodes important for shaping the modern understanding of naturalness. In particular, aesthetic arguments were absent from my discussion. Famously, Dirac believed that any large numbers must be explicable in terms of simple mathematical relations between parameters of $\mathcal{O}(1)$, a requirement that he motivated on aesthetic grounds \cite{dirac1937}. I have omitted discussion of this for two reasons. The first is because Dirac's concern was about large numbers in general, and made no mention of uniquely problematic features of elementary scalars; in this sense, there is a more direct connection between the observations of Weisskopf and Wilson and the later conception of naturalness. The second is because the core of my argument focuses on the construction and evolution of quantitative measures of naturalness, and in that process the aesthetic motivation for naturalness played a less central role than the physical arguments described above.  

Second, my aim has not been to suggest that Weisskopf, Wilson, or others who noted that elementary scalar particle masses had uniquely problematic properties, prior to the work of Susskind \cite{Susskind1979} and 't Hooft \cite{hooft1980}, were operating with the later, fully developed conception of naturalness that came to be associated with those properties. My aim has rather been to present evidence that the later conception of naturalness, and subsequent quantifications of that notion, emerged from a proximate pre-history of discussions in which elementary scalars were recognized to be uniquely problematic on primarily physical grounds and not, for example, solely or even primarily based on aesthetics. In particular, my aim has been to demonstrate the following:

\begin{enumerate}
\item The mass terms of elementary scalar particles have been recognized as uniquely problematic on the basis of physical arguments since at least the late 1930s. 

\vspace{\baselineskip}

\item Two of the problematic features now commonly invoked to explain what is unnatural about elementary scalar mass terms -- quadratic sensitivity to the cutoff scale and failure to be protected by any symmetry -- were identified long prior to any explicit concept of naturalness.

\vspace{\baselineskip}

\item The identification of these features was immediately put to use in predicting new physics and motivating theory selection, roles similar to those that the concept of naturalness has played in more contemporary discussions. Weisskopf argued that the quadratic sensitivity of an elementary scalar mass could be turned into an argument for a ``critical length'' at which new physics needed to appear to ensure that the self-energy contribution to the scalar mass did not become unacceptably large, and Wilson drew upon the fact that elementary scalar mass terms are not protected by any symmetry to rule out models that contained elementary scalars. 
\end{enumerate} 

\subsection{Naturalness and effective field theory}

It was not until the end of the 1970s that an explicit concept of naturalness was first introduced, albeit in two different guises, in papers by Susskind \cite{Susskind1979} and 't Hooft \cite{hooft1980}. By this time it was widely presumed in the physics community that there existed an energy scale $\Lambda$ -- a UV cutoff -- near which the Standard Model becomes inapplicable, and that physical processes at energies above $\Lambda$ are governed by a new quantum field theory; put otherwise, the physics community had come to believe that what I labeled Wilson's possibility (b) was likely realized in nature. 

Susskind introduces the concept as follows:

\begin{quote}
Aside from the subjective esthetic argument, there exists a real difficulty connected with the quadratic mass divergences which always accompany scalar fields. These divergences violate a concept of naturalness which requires the observable properties of a theory to be stable against minute variations of the fundamental parameters \cite[p. 2619]{Susskind1979}.
\end{quote}

\noindent By ``fundamental parameters'' Susskind means the couplings defined at the scale of the UV cutoff $\Lambda$. Elementary scalar particles are unnatural because their measurable, physical masses (which, for illustration, Susskind imagines to be $\sim 1$ GeV) depend very sensitively on the value of the effective, running mass at the cutoff scale (which Susskind imagines to be the Planck scale $\Lambda = 10^{19}$ GeV). 

Let me pause here for one last historical aside. Susskind attributes this particular concept of naturalness to Wilson. In light of that, it is perhaps worth noting another remark Wilson makes in \cite[pp. 1829-30]{Wilson1971}.  Wilson is describing a scenario in which a coupling $\text{g}(\lambda)$ approaches an ultraviolet stable fixed point $x_1$ as $\lambda\to\infty$. Imagine two possible values of the coupling at low energy, $\text{g}_1(\lambda = \text{m})$ and $\text{g}_2(\lambda = \text{m})$, such that the difference between $\text{g}_1$ and $\text{g}_2$ is large. Then as $\lambda\to\infty$, that difference will be suppressed (``deamplified'' in Wilson's terminology) and the difference between the two couplings for $\lambda\gg \text{m}$ will be small. Conversely, this entails that very small changes in the value of a coupling at very high energies will translate into very large changes in its value at low energies. Wilson describes this as ``a problem,'' continuing that ``physically the qualitative nature of a given amplitude should be determined by a qualitative knowledge of the physical couplings which determine that amplitude. If one has to specify a coupling constant to 1\% accuracy in order to determine the amplitude to 50\% accuracy, there is something wrong.'' Wilson's remark is made while evaluating a different a physical situation than Susskind is discussing -- in particular, Wilson is not at that point talking about elementary scalars nor assuming the existence of a UV cutoff -- but I think Wilson's claim that there is ``something wrong'' with a sensitive dependence of low-energy observables on the precise value of high-energy couplings suggests that the concept of naturalness attributed to him by Susskind in 1979 had at least begun to take shape already in 1971.

Let us now return from this historical aside to the first introductions of an explicit notion of naturalness in the late 1970s. In \cite{hooft1980} the principle introduced by 't Hooft under the label ``naturalness'' appears, at least superficially, to be a distinct notion from that introduced by Susskind. 't Hooft introduces it as ``an order-of-magnitude restriction that must hold at all energy scales $\mu$'' [p. 135] and motivates it by invoking our experience with solid state physics; in particular, the relative insensitivity of macroscopic properties of bulk matter to small variations in the parameters characterizing its microscopic constituents. He goes on:

\begin{quote}
\ldots it is unlikely that the microscopic equations contain various free parameters that are carefully adjusted by Nature to give cancelling effects such that
the macroscopic systems have some special properties. This is a philosophy which we would like to apply to the unified gauge theories: the effective interactions at a large length scale, corresponding to a low energy scale $\mu_1$, should follow from the properties at a much smaller length scale, or higher energy scale $\mu_2$, without the requirement that various different parameters at the energy scale $\mu_2$ match with an accuracy of the order of $\frac{\mu_1}{\mu_2}$. That would be unnatural\ldots We now conjecture that the following dogma should be followed:

\vspace{\baselineskip}

\emph{at any energy scale $\mu$, a physical parameter or set of physical parameters $\alpha_i(\mu)$ is allowed to be very small only if the replacement $\alpha_i(\mu) = 0$ would increase the symmetry of the system}.

\vspace{\baselineskip}

\noindent In what follows this is what we mean by naturalness \cite[pp. 135-6]{hooft1980}.
\end{quote}

At first glance, Susskind and 't Hooft seem to be describing different principles. Susskind's concern is that the self-energy contribution to the physical mass of the elementary scalar requires the effective mass at the cutoff scale to be very finely tuned; this, in turn, entails that the measurable physical mass will be unstable against ``minute variations'' of the effective mass at the cutoff scale. He says nothing about symmetry. 't Hooft, on the other hand, offers as a necessary condition for a parameter to be natural that the parameter be ``protected'' by a symmetry; he says nothing about quadratic divergences or fine-tuning.\footnote{Indeed, 't Hooft's emphasis on the role of symmetry in his notion of naturalness has led Grinbaum to suggest that ``based upon 't Hooft's definition, [naturalness] could have received a\ldots conceptual foundation similar to that of symmetry'' \cite[p. 616]{Grinbaum2012}; see also \cite{hall2008evidence} in which the ``conventional'' approach to naturalness is described akin to a symmetry principle. I do not think this kind of justification for naturalness is compelling, but these remarks illustrate the extent to which 't Hooft's notion of naturalness has become identified with his remarks about symmetries.} In what sense, if any, can these two early formulations have been motivated by a univocal underlying intuition about naturalness?

I will shortly argue that these apparently distinct formulations of naturalness can be motivated most compellingly by understanding them as capturing distinct aspects of an underlying expectation that quantum field theory respect an ``autonomy of scales'' principle. However, I want first to introduce one more common way of phrasing the content of the naturalness principle. This is the idea that a quantum field theory is natural if and only if all dimensionless parameters and ratios of parameters either are, or can be explained in terms of, parameters that are $\mathcal{O}(1)$, a notion that Wells \cite{wells2015utility} calls ``Absolute Naturalness''.\footnote{Wells offers an interesting counterfactual history that argues strict insistence on Absolute Naturalness could have grounded a series of inferential steps leading from quantum electrodynamics to the Standard Model.} For example, Zee describes the notion of naturalness in the particle physics community as tantamount to an expectation``that dimensionless ratios of parameters in our theories should be of order unity\ldots say anywhere from $10^{-2}$ or $10^{-3}$ to $10^{2}$ or $10^{3}$'' \cite[p. 419]{Zee2010}. Stated simply as a principle about the expected size of dimensionless parameters appearing in a Lagrangian, this strikes some physicists as numerology. For example, Hossenfelder has published a blog posting titled ``To understand the foundations of physics, study numerology'' that criticizes the influence of naturalness-based reasoning in high energy physics \cite{Hossenfelder2017}. Even Wilson became suspicious of the idea that dimensionless parameters in a quantum field theory have a ``natural'' size of $\mathcal{O}(1)$ toward the end of his career, stating that the claim that small scalar masses would be unnatural ``makes no sense when one becomes familiar with the history of physics. There have been a number of cases where numbers arose that were unexpectedly small or large'' \cite[p. 13]{Wilson2005Origins}. 

Setting aside for the moment how one might justify a requirement that all dimensionless parameters be of $\mathcal{O}(1)$, it also seems as distinct from Susskind and 't Hooft's formulations of naturalness as those two formulations seem from each other. This raises two questions: how can these formulations of naturalness be justified, and what is the relationship between them?

In \cite{Williams2015}, I have argued that the answers to these two questions are closely related. One can understand these apparently distinct notions of naturalness as highlighting distinct ways that elementary scalar masses violate an underlying ``autonomy of scales'' expectation: for physical scales $\text{E}_L$ and $\text{E}_H$ that are separated by several orders of magnitude, $\text{E}_L\ll\text{E}_H$, physical processes (couplings, observables) at $E_L$ should be relatively insensitive to precise characterizations of physical processes (couplings, observables) at $E_H$.\footnote{This understanding of naturalness is also advocated in \cite{Giudice2008,Giudice2013,Barbieri2013electroWeak,Giudice2017}.} This, in turn, offers the most compelling justification for the belief in and application of various formulations of naturalness in the physics literature: to the extent that the structure of effective field theory justifies the autonomy of scales expectation, then it justifies to the same extent the seemingly distinct formulations of naturalness. To appropriate the famous picture of Descartes: the whole of naturalness is like a tree. The roots are effective field theory, the trunk is the autonomy of scales, and the branches emerging from the trunk are the precisifications, which may be reduced to three principal ones, namely quadratic divergences, symmetry, and dimensionless parameters of $\mathcal{O}(1)$. 

One can motivate the three formulations of naturalness discussed above by invoking the autonomy of scales as follows.\footnote{What follows is not primarily aimed at historically accurate exigesis: whether Susskind or 't Hooft, or others actually were motivated by an autonomy of scales expectation is not of primary importance here. That said, I think there is fairly good evidence that they were, some of which will be briefly presented here.} Susskind's formulation of naturalness as requiring ``the observable properties of a theory to be stable against minute variations of the fundamental parameters'' is straightforward to understand as formalizing an autonomy of scales expectation: indeed, given that he formulates the naturalness principle in a framework that assumes a ``fundamental'' scale $\Lambda$ and a ``light'' scale $\text{E}_L$ with $\text{E}_L\ll\Lambda$, Susskind's formulation of naturalness essentially \emph{is} what I described as the autonomy of scales expectation. 

The relationship between 't Hooft's formulation and an autonomy of scales expectation is \emph{prima facie} murkier, but examining 't Hooft's own stated motivation for introducing his notion of naturalness is clarifying. As noted above, 't Hooft begins by stating that our experience with theories of solid state physics suggests it is unlikely that the macroscopic properties of bulk matter depend sensitively on relationships between parameters in the theory governing that matter's microscopic constituents. He then states that this feature of solid state physics should be satisfied by effective field theories.\footnote{'t Hooft assumes that all the gauge theories he is investigating in \cite{hooft1980} have a UV cutoff, which he refers to as the ``Naturalness Mass Breakdown Scale'' and estimates to be at about 1 TeV.} He translates the solid state physics intuition into the quantum field theoretic framework as a requirement that the physical processes (couplings, observables) in a theory characterizing physics at the scale $\text{E}_L$ should not depend sensitively on relationships between parameters defined at a much higher scale $\Lambda$; the parameters $\alpha_i(\Lambda)$ in the effective field theory characterizing physical processes at $\Lambda$ should not have to be carefully set to cancel with one another in order for the theory characterizing physics at $\text{E}_L\ll\Lambda$ to be empirically adequate. The rough degree to which such a cancellation would be unnatural is given by the ratio $\text{E}_L/\Lambda$, and thus is determined by how many orders of magnitude separate the scales in question. 

It is this kind of delicate cancellation between properties of effective field theories characterizing physics at widely separated scales that is unnatural according to 't Hooft. He then imposes the symmetry requirement as a necessary condition for ensuring that an effective field theory not contain any parameters that require such unnatural cancellations between different scales.\footnote{A referee suggests that 't Hooft's claim that the smallness of a parameter must be accounted for with a symmetry was motivated by reflection on the central methodological role that symmetries occupy in constructing quantum field theories. I think this would have been a strong argument for 't Hooft to have given, but I am unable to find textual support for it in \cite{hooft1980}. However, I think it is an interesting and plausible conjecture about 't Hooft's attitude toward quantum field theories at the time and I mention it here as an invitation for interested parties to take up the question.}


Finally, the claim that a parameter in an effective field theory is natural if and only if it is of $\mathcal{O}(1)$ can be motivated by appeal to the autonomy of scales as follows. Consider an effective field theory of a real scalar field defined at the UV cutoff scale $\Lambda$:

$$\mathcal{S} = \int \text{d}^4\text{x} \, \frac{1}{2}(\partial_{\mu}\phi)^2 + \sum_{\text{n}\geq 2} \text{g}_n\mathcal{O}_n$$

\noindent where the operators $\mathcal{O}_n$ have mass dimension $n$ and are products of scalar field operators and their derivatives. To ensure that the action $\mathcal{S}$ has the appropriate units, the couplings $\text{g}_n$ have the form $\text{a}_n\Lambda^{-(n-4)}$ where $\text{a}_n$ is a dimensionless number. Dimensional analysis allows one to estimate, to first order in perturbation theory, the contribution of any operator $\mathcal{O}_n$ to a scattering amplitude for particles with external momenta $\text{E}_L$: 

$$\int \text{d}^4\text{x} \, \text{g}_n\mathcal{O}_n \sim \text{a}_n \left(\frac{\text{E}_L}{\Lambda}\right)^{(\text{n}-4)}$$

\noindent Once the external momenta of the particles and the UV cutoff scale are specified, one has a qualitative picture of the dominant interactions determining the scattering amplitude at a given scale $\text{E}_L$, along with a rough quantitative estimate. In particular, one can see that interactions $\mathcal{O}_n$ with mass dimension $n > 4$ will be heavily suppressed at energies $\text{E}_L\ll\Lambda$: these operators become important only for characterizing physical processes at high energies and their contributions to low-energy processes can be ignored. On the other hand, interactions with mass dimension $n < 4$ contribute more strongly at low energies than at high energies; they are important for characterizing low-energy physical processes, but become unimportant at very high energies. Throughout many areas of physics this type of dimensional analysis argument is ubiquitous, and is typically reliable.\footnote{See \cite{dine2015stress} for a complementary discussion of the way that dimensional analysis informs intuitions about naturalness.}

These dimensional analysis estimates of the contributions of interactions at different scales all depend on the dimensionless numbers $\text{a}_n$ being roughly of $\mathcal{O}(1)$. If those parameters are allowed to be very small or very large, then the discussion above is unsound: the sensitivity of scattering amplitudes at $\text{E}_L\ll \Lambda$ to, say, the operator $\mathcal{O}_{n = 8}$ can't be reliably estimated based solely on the scales involved if $\text{a}_{n=8}$ is allowed to be of $\mathcal{O}(10^{10})$. The expectation that dimensionless parameters in a quantum field theory should be of $\mathcal{O}(1)$ can thus be justified on autonomy of scales grounds: one ought to be able to give a qualitatively accurate characterization of the interactions on which a scattering amplitude will depend based solely on the scales involved in the problem.

In response to the question about the relationship between these apparently distinct formulations of naturalness, then, I claim that they can all be understood as capturing different manifestations of an underlying expectation of the autonomy of scales in effective field theory. Turning to the second question: understood as a principle about the autonomy of scales, how well motivated is naturalness as a guide to BSM model construction? 

I think that understood in this fashion there is -- or at least, there was prior to recent LHC results -- a reasonably good, but defeasible, motivation for employing naturalness as a guide when constructing BSM models. The motivation stemmed in part from structural features of effective field theory and in part from induction on our experience with naturalness-based reasoning in particle physics in the 20th century. The structural features that I have in mind are the the applicability of the Appelquist-Carazzone decoupling theorem and, more generally, the applicability of renormalization group methods. The experiences with naturalness-based reasoning in particle physics that I have in mind are (i) all couplings in the Standard Model are natural except for the mass of the Higgs boson, and (ii) in several important episodes in 20th century particle physics, successful predictions either could have been made or were in fact made based on expectations of the autonomy of scales similar to those motivating naturalness requirements.\footnote{The claim that all couplings of the Standard Model except the Higgs boson mass are natural is false if one is inclined to consider the cosmological constant as a Standard Model coupling.} The particular episodes I have in mind are the prediction of the positron, the prediction of the mass of the $\rho$-meson, and the prediction of the mass of the charm quark. Since these are discussed in some detail by others \cite{Giudice2008,Murayama2000Supersymmetry,Williams2015}, I will focus on the motivations coming from the structure of effective field theory.

Renormalization group methods, at their core, are methods for evaluating a physical process involving many scales by systematically analyzing the process one scale at a time. Suppose one has a quantum field theory, given by an action $\mathcal{S}_\Lambda$ and defined up to a chosen UV cutoff $\Lambda$, and they are interested in studying physical processes at energies $\text{E}_L\ll\Lambda$. The theory defined up to $\Lambda$ may contain degrees of freedom that are irrelevant for characterizing processes at $\text{E}_L$, and it is typically the case that calculations are more complicated in the theory defined up to $\Lambda$. Renormalization group methods allow us to separate the scales of the problem into ``momentum slices'' of width $\dd\Lambda$ and systematically analyze the problem one ``slice'' at a time. 

This is done by evaluating the functional integral associated with $\mathcal{S}_\Lambda$ between $\Lambda$ and $\Lambda - \dd\Lambda$. As one iterates this process one may encounter thresholds, such as the masses $\mu$ of heavy particles, at which some set of fields is ``integrated out'' of the theory entirely, producing a new effective theory containing only ``light'' degrees of freedom and applicable only up to a new cutoff energy $\mu$, with $\text{E}_L\ll\mu < \Lambda$. Renormalization group methods tell us that, in general, the high energy degrees of freedom that have been integrated out contribute to the resulting low-energy effective theory through (i) modifying the values of the couplings in the low-energy theory and (ii) small corrections to scattering amplitudes calculated in the low-energy theory for particles with external momenta $\ll\mu$. The widespread applicability of renormalization group methods justifies a general expectation that the characterization of physical processes at low energies $\text{E}_L\ll\Lambda$ will not depend sensitively on the structure of the theory at the cutoff scale $\Lambda$. 

A related result sometimes invoked to justify the expectation of the autonomy of scales in effective field theory is the decoupling theorem, which states precise conditions under which one can integrate heavy fields out of a functional integral and produce a consistent, low-energy effective field theory which, as above, will have altered couplings and small corrections to scattering amplitudes calculated in the low-energy theory \cite{Appelquist1975Infrared,Manoukian1983,Collins1984,Landsman1989}. Dawson, for example, has said that the problem with the Higgs boson mass stems from a failure to obey the decoupling theorem: ``The Higgs boson mass diverges quadratically! The Higgs boson thus does not obey the decoupling theorem and this quadratic divergence appears independent of the mass of the Higgs boson'' \cite{Dawson1999Introduction}. As I will discuss below, Dawson's statement is, as a technical matter, not correct but illustrates the idea that the problem with elementary scalar particle masses is that they seem to violate the \emph{spirit}, if not the letter, of the structural features that underwrite the autonomy of scales in effective field theories.

Renormalization group methods and the decoupling theorem certainly do license \emph{some} expectation that physical processes at low energies will be largely insensitive to the detailed structure of the effective field theory at much higher scales. The question, however, is whether they license the right kind of expectation: that the values of parameters in the low energy theory (and thus any observables that are functions of those parameters) will be insensitive to the values of parameters at the cutoff scale $\Lambda$. Put another way, the question is whether, and in what sense, Giudice's claim that ``[naturalness] is the consequence of a reasonable criterion that assumes the lack of special conspiracies between phenomena occurring at very different length scales. It is deeply rooted in our description of the physical world in terms of effective theories'' \cite[p. 3]{Giudice2013} can be justified by renormalization group methods and the decoupling theorem.

The short answer is that requiring that a quantum field theory be natural demands a more stringent autonomy of scales than we are strictly licensed to expect by these structural features of effective field theory. Nothing in the structure of effective field theory just described places any restrictions on how sensitive observables calculated in a low-energy effective field theory, obtained by integrating out high-energy degrees of freedom, can be to the values of parameters in the original effective field theory defined up to the cutoff scale $\Lambda$. This is why Dawson's statement about the decoupling theorem was not correct; the decoupling theorem does not place any constraints on the magnitude of the corrections that couplings in the low-energy effective theory can receive.\footnote{In \cite[chapter 8]{Collins1984}, for example, a pedagogical proof of the decoupling theorem is given using a scalar field theory in which the couplings in the low-energy effective theory are quadratically sensitive to the high-energy physics that is integrated out.}  Renormalization group methods and the decoupling theorem tell us only that the effects of the heavy degrees of freedom that have been integrated out can be incorporated into the couplings in the low-energy effective theory; it offers no assurances that one won't have to arrange for ``conspiracies'' between parameters at the cutoff scale $\Lambda$ to ensure that observables in the low-energy effective theory can be calculated accurately. Strictly speaking, the sense of the autonomy of scales that is underwritten by the structure of effective field theories is independent of the stricter notion demanded by naturalness.

One can read this observation as a first caveat to the idea that naturalness, understood as a principle about the autonomy of scales, is well-motivated within effective field theory. Naturalness in this sense is motivated insofar as structural features of effective field theory lead us to expect a general insensitivity of low-energy observables to the structure of the theory at much higher energies, but it goes beyond what is strictly licensed by those structural features. As I said above, unnatural theories seem to violate the spirit, though not the letter, of the relationship between scales in effective field theory.

There is a second caveat to this way of understanding naturalness. It is not a precise criterion, but rather a rough physical heuristic. How sensitive can a low-energy observable be to the values of couplings in the high-energy theory before it counts as problematic? How many orders of magnitude have to separate the scales $E_L$ and $\Lambda$ before matching of $\mathcal{O}(\frac{E_L}{\Lambda})$ between parameters at the scale $\Lambda$ becomes unnatural? How large or small can a dimensionless parameter be before it is not considered of $\mathcal{O}(1)$?\footnote{Recall Zee's somewhat relaxed attitude on this score; see also \cite[p. 103]{wells2015utility} for a similarly relaxed attitude.} I will turn attention to formal measures of naturalness momentarily, and the imprecision here may initially seem like a drawback when compared to the apparent precision of the quantitative measures of naturalness one finds in the physics literature. However, it has proven very difficult to construct any particular quantitative measure of naturalness that earns widespread approval within the particle physics community. It is a virtue of understanding naturalness as a rough physical heuristic about the autonomy of scales that it offers an explanation for the difficulty of building a satisfactory quantitative measure of naturalness. The endeavor seeks to impose unwarranted precision on a essentially imprecise concept, running afoul of an old Aristotelian dictum: ``it is the mark of an educated man to look for precision in each class of things just so far as the nature of the subject admits.''

\section{Naturalness quantified}

In order for naturalness to serve as a guide to constructing models of BSM physics and extracting predictions from them -- the use to which the principle was put already by Susskind \cite{Susskind1979} and 't Hooft \cite{hooft1980} -- a desire to formulate a more quantitative statement of the principle is understandable.\footnote{The reader is encouraged to see Grinbaum \cite[section 3]{Grinbaum2012} for a complementary discussion of the evolution of quantitative measures of naturalness and how the concept of naturalness itself underwent modifications through this process.} The first, and most influential, quantitative measure of naturalness was given by Barbieri and Giudice \cite{Barbieri1998Upper}. They explicitly aimed to quantify a notion of naturalness according to which a parameter's unnaturalness is captured by the degree to which it violates an autonomy of scales expectation: ``let us finally spend a word on the significance of the `naturalness' criterion that we are employing. The problem of the quadratic divergences of the Higgs squared mass is a serious one. There is no known example of cancellation between a quadratic divergence in the low energy theory and contributions from shorter distances'' [p. 73]. To quantify this notion, they introduce a parameter $\Delta_{i \, (\text{BG})}$ capturing the sensitivity of a low-energy observable $\mathcal{M}(\alpha_i)$ to infinitesimal variations in parameters $\alpha_i$ defined at higher energies:\footnote{In the original example, the low-energy observable is $M_Z$, standing in for the scale of EWSB, and they consider variations of parameters $\alpha_i$ related to the scale at which supersymmetry is broken in a given model. The prefactor $\frac{\alpha_i}{\mathcal{M}}$ is included to remove an overall dependence on the scale of $\alpha_i$ and $\mathcal{M}$.}

$$\Delta_{i \, (\text{BG})} = \left|\frac{\alpha_i}{\mathcal{M}(\alpha_i)}\frac{\partial \mathcal{M}(\alpha_i)}{\partial\alpha_i}\right|$$

\noindent Barbieri and Giudice then impose the requirement that a theory be considered natural if and only if $\Delta_{\text{BG}} \equiv \max\left\{\Delta_{i \, (\text{BG})}\right\} < 10$, and use this to extract upper bounds for the masses of particles in the Minimal Supersymmetric Standard Model (MSSM). 

Since \cite{Barbieri1998Upper}, there have been a number of proposals to revise this measure, the allowed value of $\Delta$, or both.\footnote{See, for example, \cite{decarlos1993oneloop,anderson1995measures,anderson1995naturalness,anderson1997Naturalness,ciafaloni1997naturalness,Athron2007New}.} My claim is that as these quantitative measures of naturalness developed, a new notion of naturalness developed too. According to this alternative notion, naturalness is a statistical property: a parameter (or model) is natural if and only if it is ``likely'' or ``probable'' according to some measure defined over some space of parameters (or models). 

This alternative notion was not introduced fully in a single paper, but developed gradually. The earliest outlines of this alternative notion begin taking shape in the early 1990s, beginning with de Carlos and Casas \cite{decarlos1993oneloop}. In that paper, de Carlos and Casas note that $\Delta_{\text{BG}}$ does not capture \emph{only} the local sensitivity of an observable $\mathcal{M}(\alpha_i)$ to infinitesimal variations of $\alpha_i$; it also captures any \emph{global} sensitivity in the functional dependence of $\mathcal{M}$ on \emph{any} $\alpha_i$. For example, if $\mathcal{M}$ were the mapping $\mathcal{M}: \, \alpha \mapsto \alpha^n$ with $n\gg 1$, then the Barbieri and Giudice measure yields an unacceptably large value of $\Delta_{\text{BG}}$ independent of the value $\alpha_i$. 

This global sensitivity of the Barbieri and Giudice measure was also recognized by Anderson and Casta\~{n}o \cite{anderson1995measures}. They thought that a good measure of naturalness should declare a model unnatural only if the model required fine-tuning, and the fact that global sensitivity is not a reliable indicator of fine-tuning led them to conclude that the Barbieri and Giudice measure was inadequate as a measure of naturalness. This, in turn, motivated them to introduce an explicitly statistical notion of naturalness, which they took to more reliably indicate the degree of fine-tuning required in a model. They propose as a measure of naturalness

$$\Delta_{i \, (\text{AC})} = \frac{\Delta_{i \, (\text{BG})}}{\overline{\Delta}_{i \, (\text{BG})}}$$

\noindent This is the Barbieri and Giudice measure rescaled by an ``average'' fine-tuning $\overline{\Delta}_{i \, (\text{BG})}$ over some range of the parameter(s) $\alpha_i$. In principle, this need not signal any break with the ``autonomy of scales'' understanding of naturalness that I outlined above: one might reasonably think that a \emph{global} sensitivity of a model's low-energy observables to high-energy parameters is not terribly informative, while a model whose low-energy observables exhibit different relative \emph{local} sensitivities at distinct points $\alpha_j$ and $\alpha_k$ of parameter space might be telling us something interesting about the model at those two points.

In practice, however, Anderson and Casta\~{n}o do initiate a break from this understanding of naturalness in constructing their refined measure. In particular, they are the first to explicitly link naturalness to the statistical likelihood of specific values of high-energy parameters. They state that ``we wish to systematically clarify what measures of fine tuning best quantify our intuitive notion of naturalness and how these measures should be normalized\ldots Any measure of fine tuning that quantifies naturalness can be translated into an assumption about how likely a given set of Lagrangian parameters is'' \cite[p. 302]{anderson1995measures}. 

To accomplish this, they must first assume a probability distribution $f(\alpha_i)$ over the fundamental parameters $\alpha_i$; as they recognize, the choice of any particular $f(\alpha_i)$ ``necessarily introduces an element of arbitrariness into the construction'' [p. 302].\footnote{A number of such choices have to be made in any attempt to construct a quantitative measure of naturalness, as is discussed in some detail in \cite{Feng2013Naturalness} and \cite{Craig2013SUSY}. As I said above, I think there is a plausible argument to be made that this inevitable sense that one is making arbitrary choices stems from trying to impose unwarranted mathematical precision on an imprecise physical heuristic.} In particular, they note that ``our choice of $f(\alpha_i)$ reflects our theoretical prejudice about what constitutes a natural value of the Lagrangian parameter $\alpha_i$'' [p. 302]. Once one has selected a distribution $f(\alpha_i)$, however, one can translate this into a probability distribution over observables $X$ (e.g. $M_Z$): ``In studies of naturalness, we may ask: If the fundamental Lagrangian parameters at our high energy boundary condition are distributed like $f(\alpha_i)$, how likely is a low energy observable, $X(\alpha_i)$, to be contained in an interval $u(X)\dd X$ about $X$?'' [p. 302]. Given that the experimental values of observables like $M_Z$ are known, one can then adopt the ``interval'' $u(X) = X$; if a set of fundamental parameters distributed according to a distribution $f(\alpha_i)$ make it unlikely that $X(\alpha_i) = X$, one concludes that either (i) one's selection of a ``natural'' distribution of fundamental parameters $f(\alpha_i)$ must be revised or (ii) the model is unnatural. 

It is important for Anderson and Casta\~{n}o that one be able to distinguish between \emph{sensitivity} and \emph{naturalness}. In particular, they treat naturalness as closely related to, but distinct from, the sensitivity of certain low-energy observables to infinitesimal variations in the high-energy parameters. They draw this distinction in their conclusion:

\begin{quote}
We have analyzed the prescription popularly used to measure fine tuning. This prescription is an operational implementation of Susskind's statement of Wilson's sense of naturalness, `Observable properties of a system should be stable against minute variations of the fundamental parameters.' \emph{Because this prescription is only a measure of sensitivity, we found that it is not a reliable measure of naturalness}. We then constructed a family of prescriptions which measure fine tuning more reliably. Our measure is an operational implementation of a modified version of Wilson's naturalness criterion: Observable properties of a system should not be \emph{unusually} unstable against minute variations of the fundamental parameters'' \cite[p. 307, emphasis added]{anderson1995measures}. 
\end{quote}

\noindent Anderson and Casta\~{n}o thus present themselves as breaking with the ``autonomy of scales'' understanding of naturalness and presenting a distinct, statistical notion of naturalness.\footnote{Anderson recalls that their understanding of naturalness at the time was that ``if you imagine that the fundamental (Lagrangian) parameters had some smooth probability distribution, an observable parameter would be unnatural if the measured value of that parameter was only within some characteristic range around the measured value for an unusually small part of the parameter space relative to other values'' (personal communication). This has no essential connection to a notion of interscale sensitivity.} 

As a linguistic matter, the probabilistic notion of naturalness proposed by Anderson and Casta\~{n}o is a minor modification of the autonomy of scales notion: they conclude that a natural model is one in which observables are not \emph{unusually} sensitive to the values of fundamental parameters. As a conceptual matter, however, Anderson and Casta\~{n}o's desire to construct a notion of naturalness that translated into an assumption about a probability distribution over fundamental parameters resulted in an early and important step toward the development of a statistical notion of naturalness. 

As I said above, this development occurred gradually. As summarized by Grinbaum \cite[section 3]{Grinbaum2012}, it became increasingly popular to interpret measures of naturalness as providing information about probability distributions on parameter space. For instance, several years after \cite{anderson1995measures}, one finds Ciafaloni and Strumia treating measures of naturalness as a source of probabilistic information about parameter space \cite{ciafaloni1997naturalness}, while Giusti, Romanino, and Strumia \cite{giusti1999naturalness} speak similarly of the ``naturalness probability'' of certain regions of parameter space. By the time Athron and Miller propose to ``construct a tuning measure which determines how rare or atypical certain physical scenarios are'' \cite[p. 3]{Athron2007New} they are building on over a decade of similar interpretations of naturalness. 

In subsequent years this statistical notion of naturalness has become widespread, leading to a bifurcation of naturalness into two notions which are closely related, both historically and conceptually, but essentially distinct: one notion of naturalness according to which naturalness problems are failures of an expectation about the autonomy of scales, and a second notion according to which naturalness problems stem from a parameter (or theory) being ``unlikely'' or ``improbable''. A similar distinction has been recognized by Wells \cite{Wells2005}, who distinguishes ``Principled Finetuning'' from ``Chance Finetuning'' and suggests that these two distinct notions call for different types of solutions. 

As I discussed in the Introduction of this paper, none of the BSM physics predicted by natural extensions of the Standard Model has been detected at the LHC and this has led to a rough trifurcation of attitudes about the status of naturalness in particle physics. On the one hand, many have concluded that since ``no new physics has been so far seen at LHC with $\sqrt{s} = 8$ TeV\ldots while this is not conclusive evidence\ldots it is fair to say that the most straightforward interpretation of present data is that the naturalness ideology is wrong'' \cite{farina2013modified}. This, I think, is a quite reasonable statement insofar as one understands naturalness as a notion related to the autonomy of scales. On the other hand, one finds increasingly frequent remarks that the best hope for \emph{saving} the naturalness ideology comes from embedding problems of naturalness into a new physical setting: that of a vast landscape of effective field theories. This proposal hinges on entirely divorcing naturalness from the effective field theory context and its attendant autonomy of scales-based justification and re-casting it as a conceptually independent, purely statistical notion. The result, I claim, is a notion employed in this new physical setting that is ``naturalness'' in name only.

\section{Naturalness in the multiverse}

It is increasingly common in high-energy physics -- particle physics, quantum gravity, and early universe cosmology -- to encounter the idea that our universe may be an isolated point in a much larger multitude of causally disconnected universes: a multiverse. Most commonly, the picture is that of a large number of effective field theory vacua that arise from different compactifications of extra dimensions in string theory models, with these vacua populated by some mechanism for eternal cosmic inflation.\footnote{For pedagogical discussion of the details of compactification mechanisms and the origin of the landscape, see \cite{ibanez2012stringpheno} or \cite{denef2008houches}.} The result is a large space of low-energy effective field theories -- the string landscape -- across which low-energy physics such as symmetries, parameter values, and particle content can vary. 

One also often finds it suggested that the notion of naturalness developed in the context of effective field theory in a single universe can be conservatively embedded into a multiverse setting. This trend is noted by Giudice, who remarks that 

\begin{quote}
There is already ongoing activity on how the concept of naturalness could be reshaped in post-natural times\ldots I will only comment on a single new trend: the idea that the explanation of Higgs naturalness may not lie behind some still undiscovered symmetry, but within the cosmological evolution of the universe. The most daring approach of this kind is based on a multiverse populated by eternal inflation, in conjunction with the idea that fundamental parameters\ldots [may be] dynamical variables that take different values in a landscape of vacuum states \cite[p. 8]{Giudice2017}.
\end{quote} 

\noindent It is not hard to substantiate Giudice's claim that this is a trend. Consider a representative sample of quotations: 

\begin{itemize}
\item Dine, Gorbatov, and Thomas: ``[We] stress that within the landscape, conventional notions of naturalness are sharpened, not abrogated'' \cite{dine2008low}.
\item Silverstein: ``I find the statistical program for seeking generic properties of string vacua extremely interesting, particularly in its prospects for refining our notions of naturalness'' \cite{silverstein2004counter}.
\item Carroll: ``the possible epistemological role of the multiverse is to explain why our observed parameters are natural'' \cite{Carroll2006}. 
\item Douglas: ``Moduli stabilization also determines the distribution of vacua...and thus the distribution of couplings and masses in the low energy effective theory. One can make detailed statistical analyses of this distribution, which incorporate and improve the traditional discussion of naturalness of couplings'' \cite{douglas2013string}. 
\end{itemize}

\noindent The notion of naturalness upon which all of these authors draw is the statistical one whose development was sketched above. Douglas has re-cast this statistical notion in the setting of the string landscape as a principle of \emph{stringy naturalness}: 
\begin{quote}
An effective field theory (or specific coupling, or observable) $T_1$ is \emph{more natural} in string theory than $T_2$, if the number of phenomenologically acceptable vacua leading to $T_1$ is larger than the number leading to $T_2$ \cite{douglas2004basic,douglas2007flux,douglas2013string}.
\end{quote}

\noindent Though not everyone arguing that a multiverse may offer solutions to naturalness problems uses Douglas's exact definition, the statistical notions employed are sufficiently similar that I will treat ``stringy naturalness'' as a catch-all term for the notion(s) of statistical naturalness employed in the multiverse.

In a multiverse setting, then, naturalness problems are transformed from problems concerning the autonomy of scales to problems concerning the counting of vacua with phenomenologically acceptable values of the mass of the Higgs boson and/or the cosmological constant. There are several properties of the string landscape that might immediately give one pause.\footnote{My thanks to a referee for urging me to address this.} In order to count phenomenologically acceptable string vacua, one needs both a measure for counting and a clearly specified space of phenomenologically acceptable string vacua; in a string landscape with the vacua populated by some mechanism for eternal inflation, one has neither. The lack of a non-arbitrary measure is well-known, and stems from the need for eternal inflation: as Guth \cite{guth2007eternal} points out, ``In an eternally inflating universe, anything that can happen will happen; in fact, it will happen an infinite number of times.''\footnote{See also \cite{schellekens2013life} for a discussion of this problem that is more directly focused on the string landscape.} This makes clear the problem with imagining that one can simply count vacua and then compare relative frequencies: one cannot define the relative frequency of vacua with property A to vacua with property B if both numbers are infinite. As \cite{guth2007eternal} points out, one can introduce a regularization method to get a meaningful ratio, but doing so produces results that depend sensitively on apparently arbitrary choices about the regularization; for example, if one orders the natural numbers $\mathbb{N}$ as $\left\{1, \, 3, \, 2, \, 5, \, 7, \, 4, \, \ldots \right\}$ and takes $N\to\infty$ one gets that the relative frequency of odd numbers in $\mathbb{N}$ is two-thirds, while choosing  $\left\{1, \, 2, \, 3, \, 4, \, 5, \, \ldots \right\}$ yields a relative frequency of odd numbers in $\mathbb{N}$ of one-half. 

I will argue below that by re-casting in the string landscape the statistical notion of naturalness born in BSM physics, one loses the effective field theory structure that made it seem troubling that the measured value of the Higgs mass was highly sensitive to variations of the high-energy parameters; that is, one loses the motivation for viewing naturalness problems as ``problems'' at at all. The measure problem entails that the situation is even worse than that: one also loses any mathematically well-defined, non-arbitrary notion of probability that was associated with the statistical notion of naturalness in the BSM setting. Even if, in the BSM context, one preferred the statistical conception of naturalness to the autonomy of scale notion, the mathematically well-defined notion of probability attendant to that statistical conception that contributed to its appeal in the BSM context does not carry over to the string landscape. 

In fact, for all of the trouble one faces in defining a measure on the string landscape, there is an even more fundamental question that is unanswered: what, precisely, is the space on which one seeks to define a measure? Specifically, it is unclear how much volume of the space of \emph{all} apparently consistent low-energy effective field theories is occupied by effective field theories that can arise from string theory. If the answer is that \emph{any} consistent low-energy effective field theory can be produced by some string compactification, then the space on which one is attempting to define a measure is the space of all low-energy effective field theories. This attitude is not uncommon: as Brennan, Carta, and Vafa \cite[p. 20]{brennan2017string} state, ``there
has been a distinct philosophical shift in the community over the past decade$\ldots$ Instead of starting with fully-fledged string theory and studying the compactifications down to 4D, many have started studying effective four dimensional quantum field theories$\ldots$The common lore is that because the string landscape is so large, it is likely that any consistent looking lower dimensional effective field theory coupled to gravity can arise in some way from a string theory compactification.'' 

Contrary to this common lore, it has recently been conjectured that requiring that a low-energy effective field theory be UV-completable into string theory places strong constraints on the properties that the effective field theory can have. According to this conjecture, most apparently consistent low-energy effective field theories \emph{cannot} arise from string compactifications, with the result that the string landscape occupies a very small volume in the space of all low-energy effective field theories \cite{vafa2005string,ooguri2007geometry,brennan2017string}.\footnote{See also \cite{adams2006} for an early investigation of the properties an effective field theory should satisfy if it is to have a UV completion.}   Those effective field theories that cannot be UV-completed into string theory constitute the ``swampland'' and the conjecture that the landscape occupies a very small volume in the enormous space of all low-energy effective field theories is the ``swampland conjecture.'' The truth or falsity of the swampland conjecture is relevant to any attempt to employ a statistical notion of naturalness on the string landscape; insofar as one does not know which low-energy effective field theories can arise from string theory, one does not even have a clear specification of the space on which they hope to define a measure. 

Whether the swampland conjecture is true or not, the space on which one is trying to define a naturalness measure is one in which not merely the parameter values, but also the symmetries, particle content, and so on can vary at different points in the space. This is starkly different from the situation in BSM applications of naturalness where one selects a single model, with fixed particle content and symmetry group, and evaluates the sensitivity of its observables to variations around different points in a clearly specified parameter space. I take these considerations to indicate that even if a statistical notion of naturalness could be well-defined in the string landscape, it would have little conceptual relationship to even the statistical notion of naturalness that is employed in the context of BSM physics, and would play a quite distinct methodological role from that it plays in BSM physics.

All of that said, advocates of re-casting naturalness in a multiverse often prefer to proceed as if the only feature of the multiverse that is relevant is that it is a large space of low-energy effective field theories across which the values of parameters, like the mass of an elementary scalar particle or the cosmological constant, can vary. Even setting aside the above concerns and proceeding with the discussion on these terms, it quickly becomes clear that the conception of naturalness being employed in these discussions has essentially nothing to do with the notion that was well-motivated within the structure of effective field theory; the result is that in a multiverse, one loses the traditional justification for viewing naturalness problems as ``problems'' at at all.

One preliminary indication of this is that just the numerical value of the Higgs boson mass was never the compelling problem for BSM physics; rather, that the value seemed puzzling was a symptom of its sensitivity to the details of the Standard Model's structure at high energies. According to those who developed the notion of naturalness -- Wilson, Susskind, 't Hooft, Barbieri and Giudice, and others -- it was this \emph{sensitivity} that was unnatural in the context of effective field theory. In that sense, the counting problems that stringy naturalness picks out as requiring solutions have no relationship to the notion of naturalness that is motivated within effective field theory and the problems it identifies.

It is true that stringy naturalness bears some conceptual relationship to the statistical notion of naturalness whose development was sketched above. As was argued above, that statistical notion itself has tenuous connection to the original conception of naturalness as an autonomy of scales requirement, but it is certainly true that the statistical conception of naturalness retains at least some connection to the autonomy of scales notion within the context of effective field theory. If for no other reason, this is ensured by the fact that the quantitative measures constructed by Anderson and Casta\~{n}o or Athron and Miller, for instance, are variations on the Barbieri and Giudice measure, which was explicitly constructed to track the sensitivity of low-energy observables to variations of high-energy parameters. Insofar as one is supposed to measure stringy naturalness by something like a straightforward counting of vacua, it does not enjoy even this formal relationship to the Barbieri and Giudice measure. Whatever conceptual relation there is between the autonomy of scales notion of naturalness and the statistical notion of naturalness in the BSM context, it is entirely severed when the latter is re-cast as ``stringy naturalness'' in a multiverse context.

There is a specific wedge issue that is helpful for seeing the way in which the autonomy of scales notion of naturalness and stringy naturalness come apart: whether low-energy supersymmetry should be considered natural. Low-energy supersymmetry is, of course, paradigmatically natural according to the autonomy of scales conception of naturalness; its ability to provide an extension of the Standard Model that naturally explained the stability the EWSB has long been one of the most popular theoretical arguments in its favor. By contrast, and somewhat remarkably, low-energy supersymmetry may count as \emph{unnatural} according to stringy naturalness! 

It has long been unclear whether a low-energy scale of supersymmetry breaking is statistically favored among the effective field theories in the landscape. Attempts to analyze the statistical distribution of low-energy effective field theories in the landscape have led to conflicting results; see \cite{douglas2003statistics,denef2004distributions,susskind2005supersymmetry,douglas2007flux,dine2008low,douglas2013string} and many references therein. This, combined with the definition of stringy naturalness as ``statistically favored in a multiverse'' is what allows Douglas \cite{douglas2013string} to claim that an argument ``from stringy naturalness'' suggests that string theory prefers a supersymmetry breaking scale of $30-100$ TeV. That those who employ stringy naturalness are not working with the same notion of naturalness as Wilson, Susskind, 't Hooft, or Barbieri and Giudice is clear from the fact that this is a supersymmetry-breaking scale several orders of magnitude too high for the MSSM and other simple supersymmetric extensions of the Standard Model to be natural, according to the autonomy of scales notion of naturalness. 

I certainly do not claim to have any particular insight into whether low-scale supersymmetry breaking is, in fact, statistically favored in the string landscape, but such insight is unnecessary for my present aim.\footnote{For what it is worth, the attitude expressed by Arkani-Hamed and Dimopoulos\cite{ArkaniHamed2005} seems to me quite reasonable: ``One might think that low-energy SUSY with $m_S\sim \text{TeV}$ is preferred, since this does not entail a large fine-tuning to keep the Higgs light. However, this conclusion is unwarranted$\ldots$ without a much better understanding of the structure of the landscape, we can't decide whether low-energy SUSY breaking is preferred to SUSY broken at much higher energies.''} I want only to show how thoroughly divorced the notion of stringy naturalness employed in a multiverse setting is from the autonomy of scales notion of naturalness developed in the context of effective field theory. All that is needed to achieve that goal is the fact that it is widely considered an open question whether low-energy supersymmetry is stringy natural in a multiverse setting, while low-energy supersymmetry is the \emph{paradigmatically} natural scenario in a effective field theory context. The fact that an effective field theory with low-energy supersymmetry breaking can be considered natural according to the autonomy of scales notion of naturalness while that very same model, if embedded in a multiverse, could simultaneously be deemed \emph{stringy-unnatural} because there are insufficiently many \emph{other} vacua with low-energy supersymmetry breaking, demonstrates clearly that there are two independent notions of naturalness in play.

\section{Conclusion}

I want to conclude with two further remarks. The first is an additional comment on the way in which the two notions of naturalness come apart. On the one hand, whether an effective field theory is natural is a ``local'' property according to the autonomy of scales conception of naturalness: it is determined entirely by the sensitivity of a theory's low-energy observables to variations of fundamental parameters within a small neighborhood of a selected point in parameter space. In the Barbieri and Giudice measure, for instance, this small neighborhood is infinitesimal: one simply takes derivatives of the selected low-energy observables with respect to the selected high-energy parameters. It is a notion that tracks the stability of the theory's low-energy observables against these minute variations around selected points in parameter space. On the other hand, stringy naturalness is a thoroughly ``global'' notion: in order to determine whether a coupling, observable, or effective field theory is stringy natural, one must examine the \emph{entirety} of the string landscape of low-energy effective field theories. It is thus unsurprising that the two notions can render conflicting verdicts on any given model: they are tracking independent properties, and are determined by investigating very different volumes of two very different spaces.

One may object to this diagnosis on the grounds that while the value that the Barbieri and Giudice measure assigns to $\Delta_{i \, (\text{BG})}$ depends only on an infinitesimal neighborhood around a point  $\alpha_i$ in parameter space, this is not true even for other, non-stringy measures of naturalness.\footnote{My thanks to a referee for offering this objection.} For example, the Anderson and Casta\~{n}o measure relies on information about the degree of global sensitivity of an observable $X$ to \emph{all} parameter values $\alpha_j$ over some chosen non-infinitesimal volume of parameter space. They then incorporate this non-local information into their proposed measure of naturalness, which is essentially $\Delta_{i \, (\text{BG})}$ rescaled to eliminate the global sensitivity. The later measure proposed by Athron and Miller \cite{Athron2007New} is also non-local, roughly defined as the ratio between the volume in parameter space capable of reproducing the measured values of a model's low energy observables and the ``typical'' volume in parameter space that one would expect to reproduce those observables in that model. 

My response to this concern is as follows. The distinction I have aimed to draw is between an ``autonomy of scales'' conception of naturalness and a statistical notion of naturalness. I have argued that the definitions offered by Wilson, Susskind, and 't Hooft, and the quantitative measure introduced by Barbieri and Giudice, are best understood as motivated by the ``autonomy of scales'' conception, and that the measures proposed by Anderson and Casta\~{n}o, Athron and Miller, and advocates of stringy naturalness all, to different degrees, break from the ``autonomy of scales'' conception. The fact that these latter measures of naturalness are to some degree non-local can be seen as a symptom of the fact that they are quantifying a conception of ``naturalness'' that is distinct from the ``autonomy of scales'' conception. 

In particular, Athron and Miller state explicitly that they are breaking from the autonomy of scales conception and aim to construct a measure of statistical typicality: ``fine tuning may also be characterized by instability. It is this instability which the traditional measure is exploiting. Instead we wish to construct a tuning measure which determines how rare or atypical certain physical scenarios are'' \cite[p. 3]{Athron2007New}. The fact that the resulting measure is non-local should not be taken as evidence that the autonomy of scales conception of naturalness is itself non-local. 

Although Anderson and Casta\~{n}o also explicitly break with the autonomy of scales conception of naturalness, the close connection of the measure they propose with the Barbieri and Giudice measure makes its analysis less straightforward. I think a plausible, albeit strict, reading of the autonomy of scales conception of naturalness \emph{does} entail that global sensitivity is an informative a property of a model. This would still allow naturalness to be used in practice, just as it is now, when evaluating \emph{different} models at points in some jointly allowed region of parameter space: for instance, if a set of observables in the Standard Model are more sensitive than are those same observables in the MSSM to variations around every point $\alpha_j$ in some allowed region of parameter space, one might think that this global sensitivity should be factored into our judgment about whether the MSSM is more natural than the SM, \emph{even at a specific point} in parameter space $\alpha_k$. 

Adherence to this strict reading of the autonomy of scales conception would, however, affect how one uses naturalness measures when comparing the \emph{same} model at different points. The naturalness value of $\Delta_{i \, (\text{BG})}(\alpha_i)$ or $\Delta_{j \, (\text{BG})}(\alpha_j)$ would not be intrinsically informative, but only provide us with information about the relative naturalness of the model at different points, i.e. only $\Delta_{i \, (\text{BG})}(\alpha_i) - \Delta_{j \, (\text{BG})}(\alpha_j)$ would be meaningful. On this reading of the autonomy of scales conception of naturalness, one maintains a local measure of naturalness at the expense of accepting that it is only comparatively meaningful. The belief that measures of naturalness have, at best, comparative meaning but no intrinsic meaning is not unique; for example, after a review of some of the apparently arbitrary decisions that one has to make when constructing a quantitative measure of naturalness, Craig \cite[p. 7]{Craig2013SUSY} concludes that ``it is clear that measures of tuning have no intrinsic meaning. They may have some comparative value in terms of contrasting models, but even this is not absolute.'' By contrast, Anderson and Casta\~{n}o aim to construct a measure of naturalness that is intrinsically meaningful, but at the expense of making the notion explicitly non-local. 

I argued above that Anderson and Casta\~{n}o initiate a break with the autonomy of scales conception of naturalness, but this is due to their tying naturalness to a probability distribution over fundamental parameters, not by introducing a non-local rescaling of what is essentially the Barbieri and Giudice measure. In this particular case, choosing between a local, comparatively meaningful measure or a non-local, intrinsically meaningful measure of naturalness strikes me as akin to choice of convention with little conceptual significance. I hasten to add that this is not the case for the stringy naturalness; in that case, there is no alternative, local way to construe the ``global'' notion of stringy naturalness employed in the landscape. 

My second remark concerns the purported ability of the multiverse to solve naturalness problems; specifically, the manner in which this purported ability has been presented as supporting evidence for the existence of a multiverse, and thus as supporting evidence for theories that evidently give rise to a multiverse. Hall and Nomura, for example, claim that ``evidence for the multiverse can be found in three different arenas: the cosmological constant, nuclear physics, and electroweak symmetry breaking. In all three cases, the conventional approach$\ldots$ leads to naturalness problems$\ldots$ In each arena the multiverse easily and generically solves the naturalness problem'' \cite[p. 39]{hall2008evidence}, a claim which they base on a statistical notion of naturalness much like the stringy notion discussed above. In a similar, though more restrained, spirit, Douglas \cite{douglas2004basic} writes ``We only live in one vacuum. However$\ldots$ vacuum multiplicity can help in solving the cosmological constant problem$\ldots$ In the absence of other candidate solutions to the problem, we might even turn this around and call these ideas evidence for the hypothesis that we are in a compactification with many hidden sectors.'' 

Problems of naturalness have driven much of BSM physics since the late 1970s and it would certainly count as an accomplishment of a multiverse theory if it were able to solve those problems. However, as we have seen, the claim that a multiverse can solve the naturalness problem(s) in the Standard Model trades on a notion of ``naturalness problem'' that has fundamentally no conceptual overlap with the naturalness problem(s) that 40 years of BSM theorizing have aimed to solve. Those who present as evidence for a multiverse the claim that it can solve the naturalness problem(s) of the Standard Model are equivocating between two essentially different notions of naturalness. Thus, while statistical analyses of the string landscape are certainly of great interest, and it would be an important discovery to determine that string theory predicts that ``most'' low-energy effective field theories in a multiverse contain an elementary Higgs boson with a mass of 125 GeV or a very small cosmological constant, it is misleading to present this as offering a natural solution to these problems: such a statistical analysis could not and does not provide a solution to the naturalness problem(s) that particle physics has aimed at solving since the 1970s. 

\section*{Acknowledgments}
I would like to thank Tony Duncan, Michael Miller, and an insightful referee for this journal for helpful comments on an earlier draft, and Greg Anderson and Diego Casta\~{n}o for helpful correspondence about the motivation for the notion of naturalness they introduced in \cite{anderson1995measures}. I would also like to thank audiences at the Aachen workshop ``Naturalness, Hierarchy, and Fine-tuning,'' the University of Michigan workshop ``Foundations of Modern Physics: the Standard Model after the Discovery of the Higgs Boson,'' and at Balliol College, Oxford for their valuable feedback.

\end{document}